\documentclass[aps,preprint,groupedaddress,superscriptaddress,showpacs]{revtex4-1}

\usepackage{graphicx}
\usepackage{color}
\usepackage{amsmath}
\usepackage{amsfonts}
\usepackage{amssymb}
\usepackage{dcolumn}
\usepackage{hyperref}

\begin{document}
  \title{Gender gap in the ERASMUS mobility program}

 \author{L. B\"{o}ttcher}
  \affiliation{Computational Physics for Engineering Materials, IfB, ETH Zurich, Wolfgang-Pauli-Strasse 27, CH-8093 Zurich, Switzerland}
\author{N. A. M. Ara\'ujo}
 \affiliation{Departamento de F\'{\i}sica, Faculdade de Ci\^{e}ncias, Universidade de Lisboa, P-1749-016 Lisboa, Portugal, and Centro de F\'isica Te\'orica e Computacional, Universidade de Lisboa, Campo Grande, P-1749-016 Lisboa, Portugal}
 \author{J. Nagler}
 \affiliation{Computational Physics for Engineering Materials, IfB, ETH Zurich, Wolfgang-Pauli-Strasse 27, CH-8093 Zurich, Switzerland}
  \author{J. F. F. Mendes}
 \affiliation{
Departamento de Fisica and I3N, Universidade de Aveiro, 3810-193 Aveiro, Portugal}
 \author{D. Helbing}
 \affiliation{Computational Social Science, ETH Zurich, Clausiusstrasse 50, CH-8092 Zurich,
Switzerland}
 \author{H. J. Herrmann}
 \affiliation{
 ETH Zurich, Wolfgang-Pauli-Strasse 27, CH-8093 Zurich,
Switzerland, and Departamento de F\'isica, Universidade
Federal do Cear\'a, 60451-970 Fortaleza, Cear\'a, Brazil}

\begin{abstract}
Studying abroad has become very popular among students. The ERASMUS
mobility program is one of the largest international student exchange
programs in the world, which has supported already more than three
million participants since 1987. We analyzed the mobility pattern within
this program in 2011-12 and found a gender gap across countries and
subject areas. Namely, for almost all participating countries, female
students are over-represented in the ERASMUS program when compared to
the entire population of tertiary students. The same tendency is
observed across different subject areas. We also found a gender
asymmetry in the geographical distribution of hosting institutions, with
a bias of male students in Scandinavian countries. However, a
detailed analysis reveals that this latter asymmetry is rather driven by
subject and consistent with the distribution of gender ratios among
subject areas.
\end{abstract}

\maketitle
 
\section*{Introduction~\label{sec:intro}}
Statistical analyses of big data sets have revealed interesting patterns 
related to human mobility. For example, from the trajectory of mobile phone 
users, it was possible to identify temporal and spacial regularity in the mobility 
patterns, with a characteristic travel distance and a small set of frequently visited locations 
for each individual~\cite{Gonzalez08}. Using data from global connectivity and 
epidemic spreading, Brockmann and Helbing could successfully predict the disease 
arrival time and/or sources for different diseases~\cite{Brockmann13}. Also, from 
the database of airports and alternative connections between these
airports it was possible to reveal a core-periphery structure in the World Airline
Network, consisting of a strongly connected core and a weakly connected, tree-like,
periphery~\cite{verma14}.  Here, we use similar tools to evaluate possible gender
differences in the mobility pattern of ERASMUS students.

ERASMUS is an European Unity exchange program that provides financial 
support to European students to study abroad. It 
 brings together more than four thousand academic
institutions and companies across $33$ countries and aims at boosting
the participants’ job prospects by encouraging international mobility
and promoting the development of personal skills, such as intercultural
awareness, openness, and flexibility~\cite{eucom2014,eucom22014,
eucom2013,parey2011,bracht06}. The participation in the ERASMUS 
program has increased impressively, from a mere three thousand 
participants in 1987 to $252827$ in
2012~\cite{eucom2014,eucom2013}. The number of participants in the
2011-12 edition corresponds to almost $1\%$ of all tertiary
students~\cite{eutertiarystat}. This impressive level of participation
makes the ERASMUS program an excellent example to study the enrollment
of students in exchange programs and to identify mobility patterns.

A comparative study of ERASMUS and non-ERASMUS students concludes that
the decision to participate is mainly affected by professional aspects
and personal preferences, although a financial barrier is also
identified~\cite{soutootero2013}. Studies of the network of ERASMUS
institutions show that the choice for a country is positively correlated
to its number of top ranked universities~\cite{gonzales2011} and that
students are typically biased towards institutions that were previously
selected by their home-university fellows~\cite{derzsi2011}. The personal 
motivation for participating in mobility programs should be interpreted in 
the context of the social environment, personal experiences, and the 
macroeconomic situation in the country of residence~\cite{vanmol2014}. 
So far, however, gender differences have not been thoroughly studied. 
The number of female students in tertiary education is definitely on the
rise~\cite{genderoecd}. According to the EUROSTAT tertiary education
statistics, the number of female students in EU-28 countries even
surpassed the number of their male fellows~\cite{eutertiarystat}.
Are the mobility patterns of male and female students the 
same or different? Here we show that, in the ERASMUS program in 2011-12,
female students are consistently over-represented, even when considering their
majority in tertiary education. This result is in sharp contrast to the
labor market, where empirical studies suggest that the mobility of
female workers is lower than the one of their male
counterparts~\cite{theodossiou09}.

\section*{Materials and Methods~\label{sec:data}}
The ERASMUS student mobility data set for the 2011-12 edition contains
the list of all participants and their home- and
host-institution/country, gender, age, nationality and subject area.
Home- and host-academic institutions are represented by their
institution code that is uniquely defined. We have the list of codes and
names for $4466$ institutions. For $1915$ of them, there is no
information available about their official name and therefore we decided
to remove them from the list. The resulting data set consists of $2551$
universities and $199488$ participants. We provide the entire data set
as \emph{Supporting Information}.

The data set also contains information about the type of mobility:
mobility for study (between two academic institutions), industrial
placement (between universities and industrial partners) or combinations
of both. In the latter case we considered only the university as the
host-institution. In the data, there are $204744$ university exchanges,
$48083$ industrial placements and only $438$ combined exchanges.

Additional statistical information about the ERASMUS program was obtained from
the statistical reports of the European Union~\cite{eucom2013}. The tertiary
education statistics in the ERASMUS countries was obtained from
EUROSTAT~\cite{eutertiarystat}.

To investigate the over-representation of female students, we first compared
the tertiary education statistics with the ERASMUS data using a null model for
which we assume that the population of ERASMUS students is randomly drawn from
the student population of the participating countries.  Both data sets are
analyzed based on the comparison of proportions of students in different
subject areas and countries. In order to better visualize the difference we
also show the ratio of both values. Our analysis is mostly based on averaging
over certain data set sub-populations.

\section*{Results~\label{sec:results}}
In 2011-12, $153468$ ERASMUS participants (about $61\%$) were female
students.  This percentage is even $1.13$ times higher than the fraction
of female students attending tertiary education in the ERASMUS
countries. This higher rate of participation is practically the same for
industrial internships and university exchanges. Note that, if
participants were randomly drawn from the entire population of $24606715$ tertiary
students, the expected number of female participants ($136527$) would differ from
the actually observed one ($153468$) more than $60$ times the standard deviation ($251$).
Thus, it is highly improbable that the over-representation of female
students is a mere statistical fluctuation. Below we analyze this gender
patterns across subject areas and countries.

\begin{figure}
\includegraphics[width=\columnwidth]{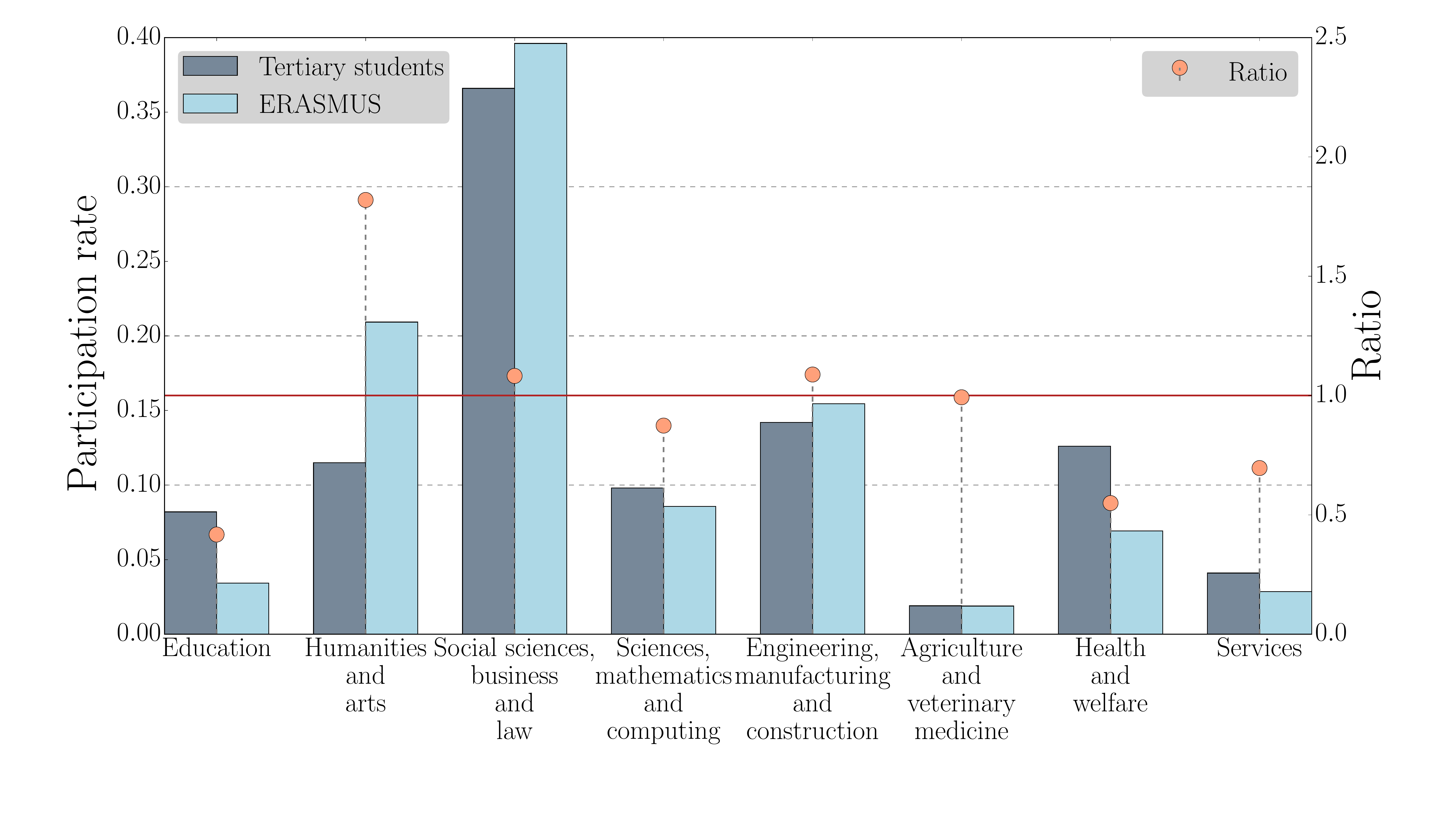}
\caption{\textbf{The participation rate in the ERASMUS program
depends on the subject area.} Participation rate in different subject
areas in the 2011-12 ERASMUS program (light blue) and in tertiary
education in the participating countries in 2011 (dark blue). The orange
circles are the ratios of these fractions.  While \textit{humanities and
arts}, \textit{social sciences, business and law}, \textit{engineering,
manufacturing and construction} are over-represented in the ERASMUS
program, the others are
under-represented.~\label{fig::barchart_subject}}
\end{figure}
\begin{figure}
\includegraphics[width=\columnwidth]{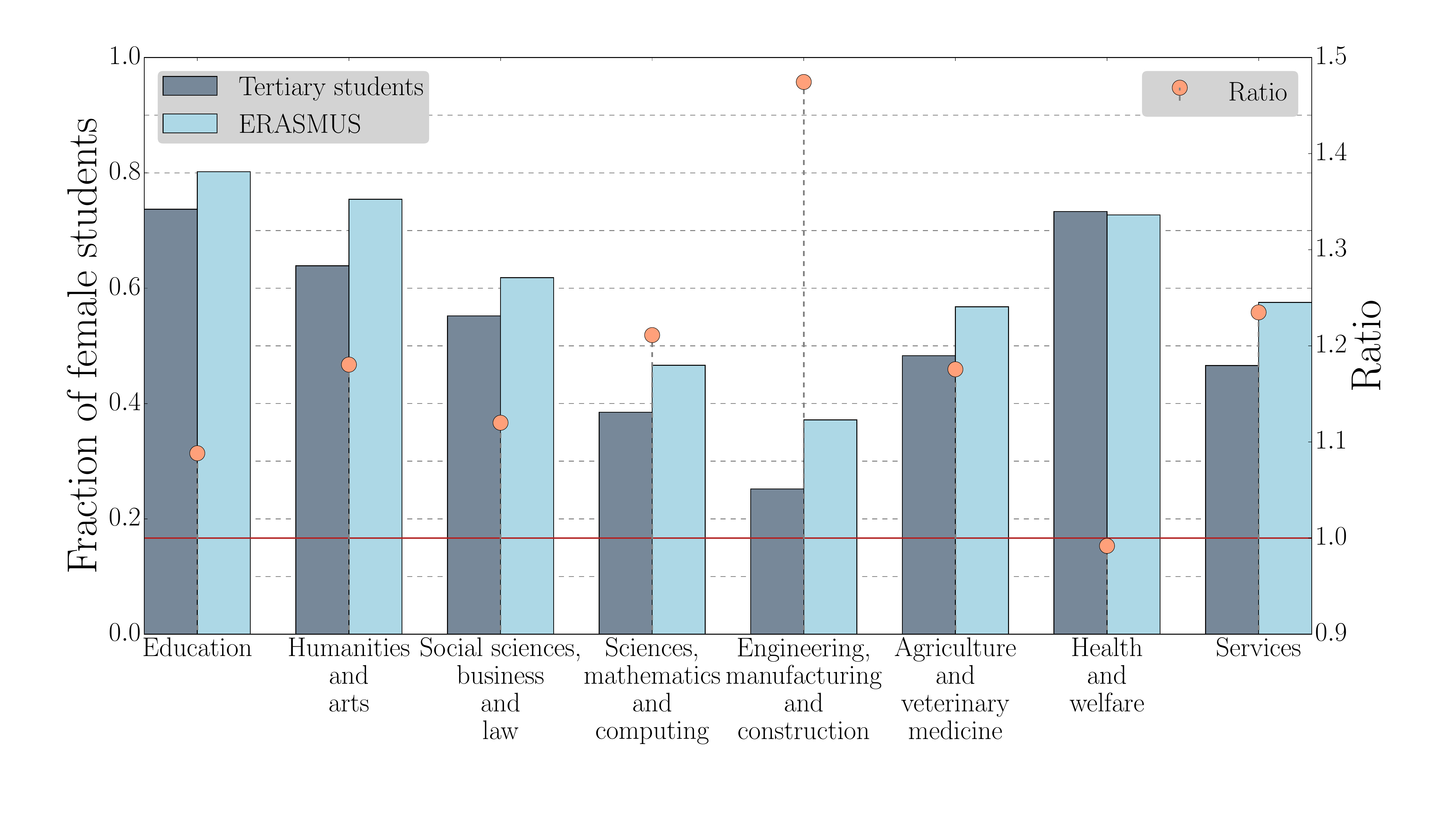}
\caption{\textbf{The over-representation of female students in the
ERASMUS program is systematic across subject areas.} Fraction of female
students in different subject areas in the 2011-12 ERASMUS program
(light blue) and in tertiary education in the participating countries in
2011 (dark blue). The orange circles are the ratios of these fractions.
For almost all subject areas, female students are over-represented in
the ERASMUS program. The only exception is \textit{health and welfare},
where the ratio is balanced.~\label{fig::barchart_female_subject}}
\end{figure}
According to the \textit{International Standard Classification of
Education} (ISCED), the population of tertiary students is divided into
eight subject areas: \textit{education}; \textit{humanities and arts};
\textit{social sciences, business and law}; \textit{sciences,
mathematics and computing}; \textit{engineering, manufacturing and
construction}; \textit{agriculture and veterinary medicine};
\textit{health and welfare}; and \textit{services}.
Fig~\ref{fig::barchart_subject} shows the participation rate
over these subject areas for the ERASMUS program and for the entire
tertiary education population in the ERASMUS countries. One sees that
certain subjects are clearly over-represented in the ERASMUS program.
For example, \textit{humanities and arts} rank second in terms of
ERASMUS participants while they rank fourth in the total population of
tertiary students. By contrast, the participation of \textit{education}
students is very low.  Fig~\ref{fig::barchart_female_subject} shows
the fractions of female students in tertiary education and ERASMUS
together with the ratio between them. In line with the agglomerated
data, females are over-represented in the ERASMUS program across all
subject areas, except for \textit{health and welfare}. For
\textit{engineering, manufacturing and construction}, an area typically
dominated by male students, the ratio is almost $1.5$ times higher.

\begin{figure}
\includegraphics[width=\columnwidth]{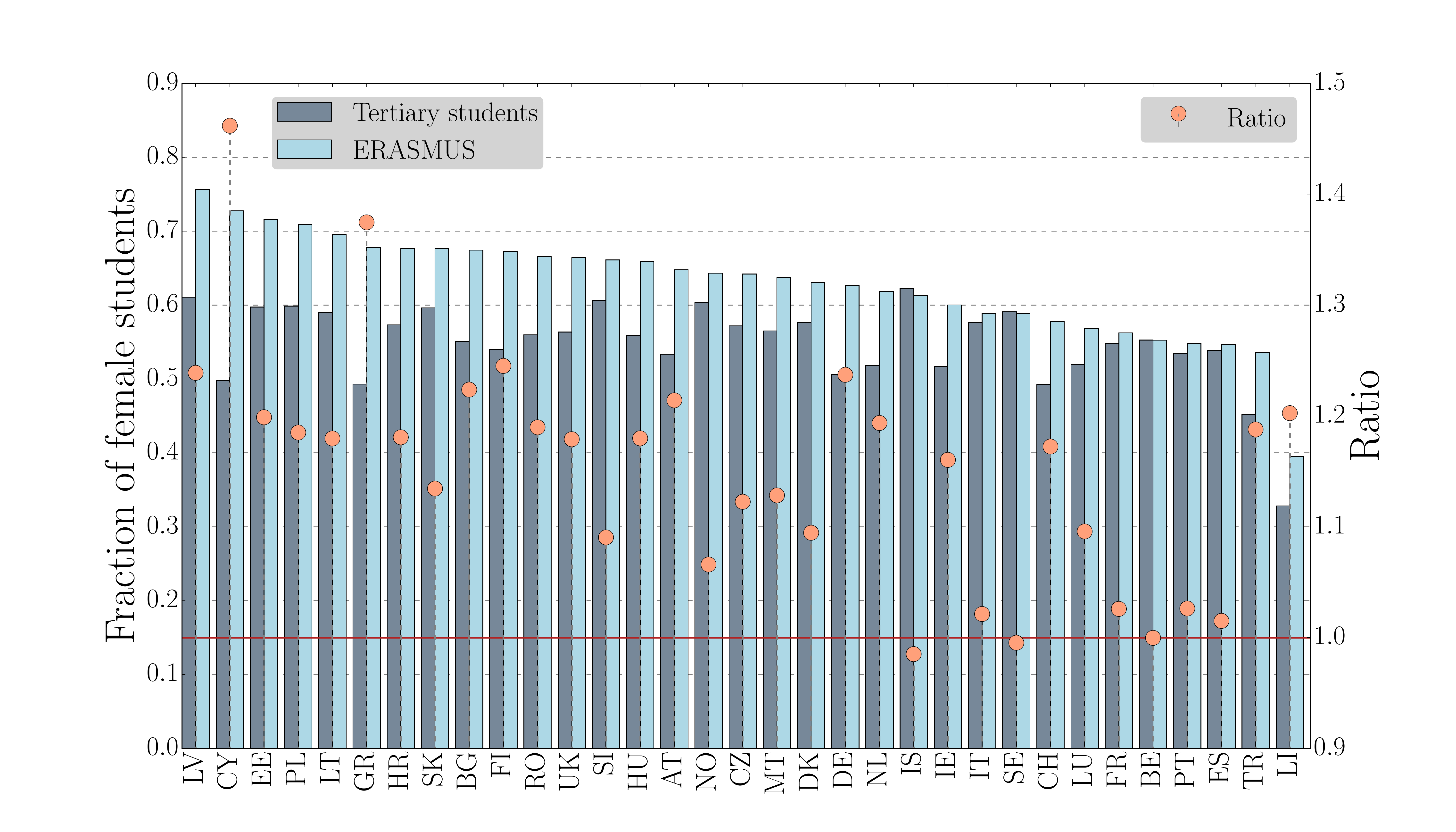}
\caption{\textbf{The over-representation of female students in the
ERASMUS program is systematic across countries.} Fraction of female
students in the 2011-12 ERASMUS program (light blue) and in tertiary
education in 2011 (dark blue) for the $33$ participating countries. The
orange circles are the ratios of these fractions. For Iceland, Italy,
Sweden, France, Belgium, Portugal, and Spain, the fractions are
similar. For all the other countries, female students are
over-represented in the ERASMUS
program in comparison to the tertiary student population.~\label{fig::barchart_female_country}}
\end{figure}
The same gender asymmetry is observed across countries.
Fig~\ref{fig::barchart_female_country} contains the fraction of
female students in the ERASMUS program, in tertiary education for all
ERASMUS countries, and the ratio between them. Only seven out of $33$
countries have a balanced fraction: Iceland, Italy, Sweden, France,
Belgium, Portugal, and Spain. For all the other countries, female
students are clearly over-represented in the ERASMUS program in comparison
to the tertiary student population. The
largest ratios are for Cyprus, Greece, Finland, Latvia, and Germany.

\begin{table}[!htp]
\caption{\textbf{Rank of sending and receiving countries.} Top five countries
with the highest number of outgoing (sending) and incoming (receiving)
students in the ERASMUS program for academic exchange, i.e., study mobility.
The fraction of female students is shown in brackets.~\label{tab::studymob_countries}}
\begin{tabular}{c c} 
\textit{Top sending} & \textit{Top receiving} \\\hline\hline
Spain: 33634 (55\%) &  Spain: 30938 (67\%) \\\hline
Germany: 27106 (62\%) & France: 22887 (69\%) \\\hline
France: 24250 (57\%) & Germany: 20885 (59\%) \\\hline
Italy: 19757 (59\%) & United Kingdom: 17697 (64\%) \\\hline
Poland: 11878 (71\%) & Italy: 17028 (65\%) \\
\end{tabular}
\end{table}

Let us now focus on the mobility between countries. For simplicity, we only
consider ERASMUS exchanges between academic institutions, disregarding
industrial placements (see \textit{Materials and Methods}). In this
way, one can keep track of institution names since they are uniquely
defined, which is very important to geographically localize them.
Table~\ref{tab::studymob_countries} contains the top five countries
sending and receiving students in the ERASMUS program, ranked by the
absolute number of ERASMUS outgoing and incoming students, respectively.
Spain ranks first in both lists. In fact, approximately $30\%$ of all
ERASMUS students are either coming from or moving to Spain. The most
popular university in the entire program is the University of Granada
(Spain) that hosts roughly $2\%$ of all ERASMUS students. If we
normalize the number of outgoing students by the total number of
tertiary students in the country, Spain ranks third ($1.7\%$ of tertiary
students), being surpassed by Luxembourg ($7.8\%$ of tertiary students)
and Liechtenstein ($3.4\%$). 

\begin{figure}
\includegraphics[width=\columnwidth]{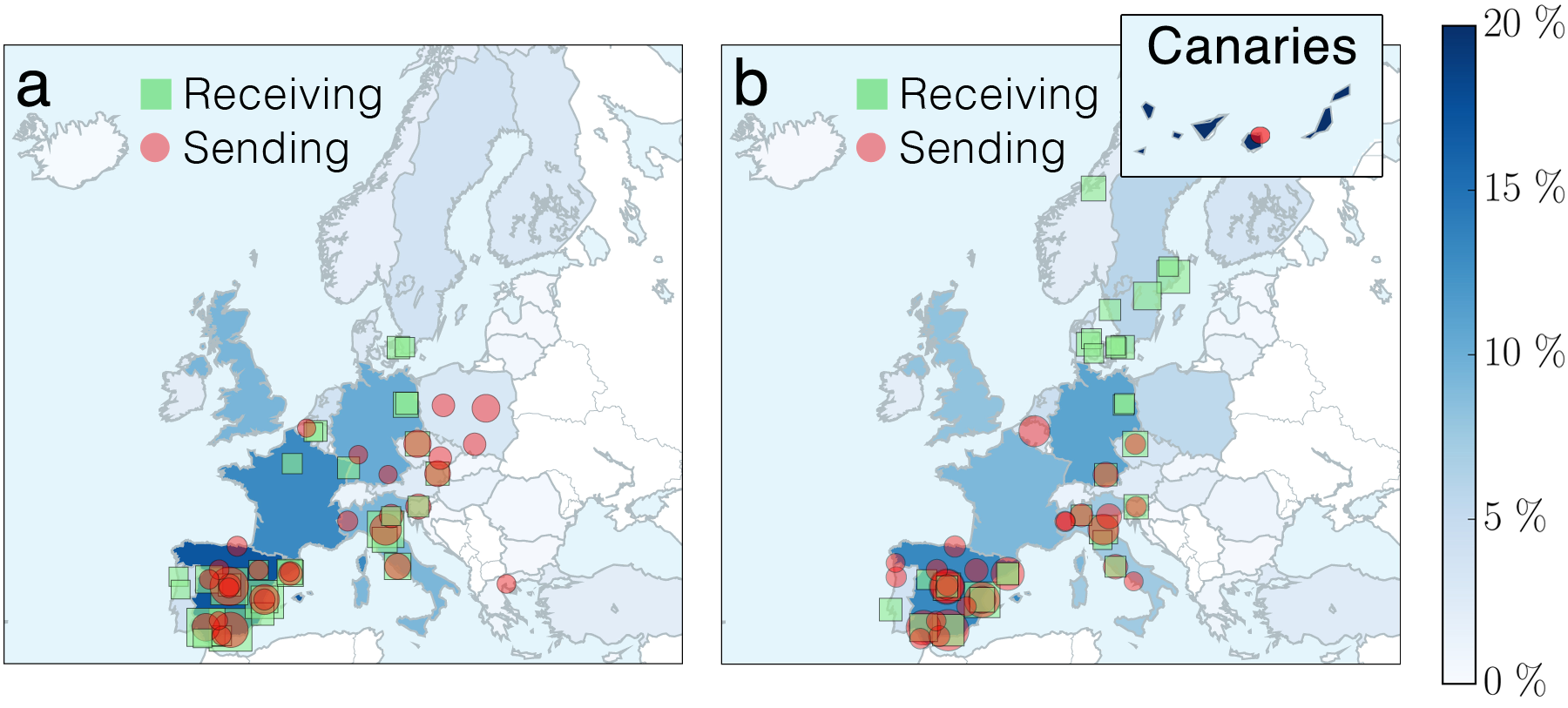}
\caption{\textbf{There is a gender asymmetry in the geographical distribution
of involved institutions.} Map of the top $30$ academic institutions ranked by
the number of outgoing and incoming (a) female and (b) male students. Red
circles represent the sending institutions and green squares the receiving
ones. The size of the symbols is proportional to the ratio of the number of
ERASMUS students to the total number of ERASMUS students in the $30$ academic
institutions. The overall fraction of receiving students in each country is
indicated by the intensity of the color of the country. The Scandinavian
universities are much more attractive to male students than to female
ones.~\label{fig::map_network_all}}
\end{figure}
\begin{figure}
\includegraphics[width=\columnwidth]{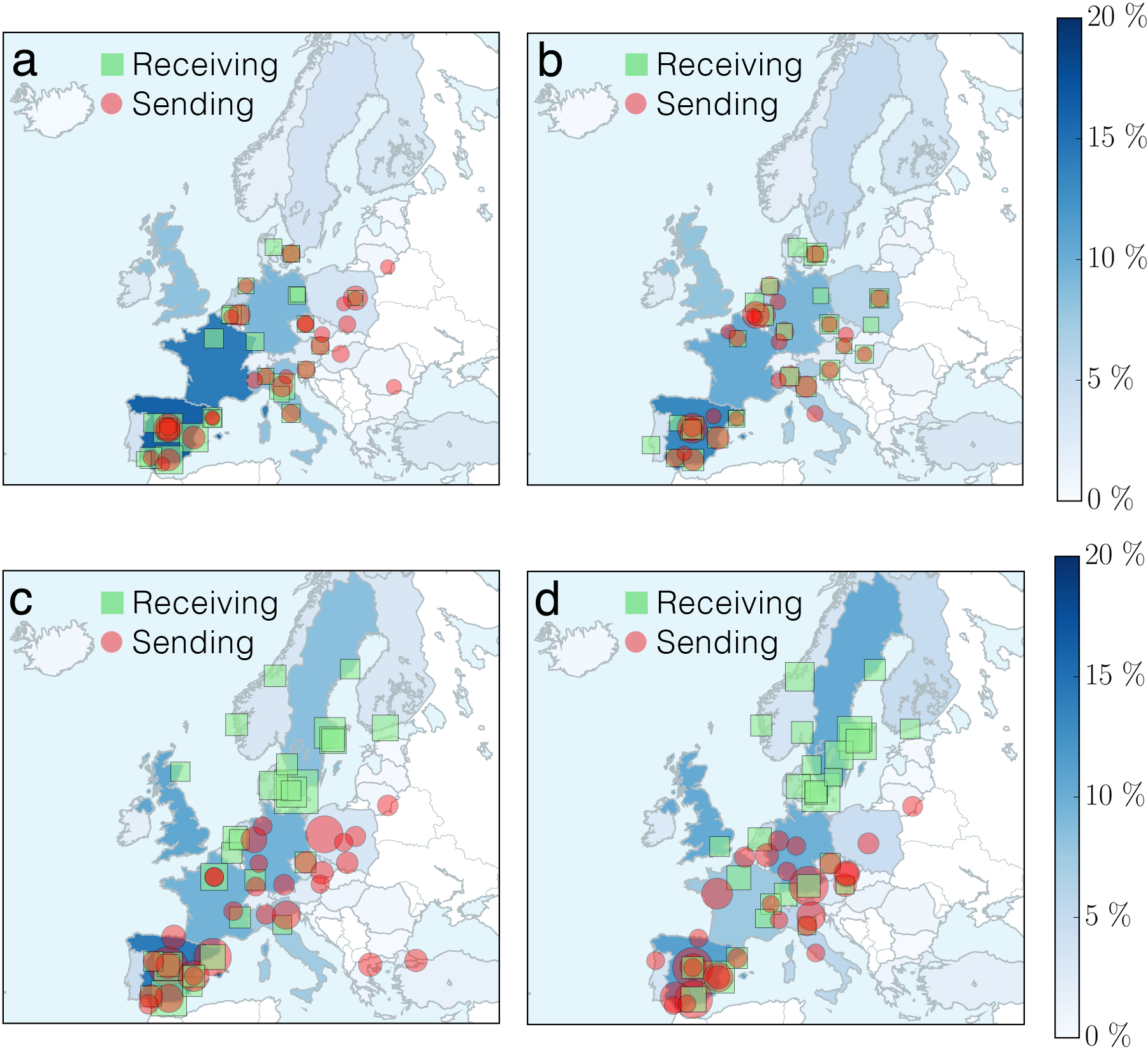}
\caption{\textbf{The geographical gender asymmetry is driven by subject area.}
Map of the top $30$ academic institutions ranked by the number of outgoing and
incoming (a,c) female and (b,d) male students, for (a,b) \textit{social
sciences, business and law} and (c,d) \textit{natural sciences, mathematics
and computing}. Red circles represent the sending institutions and green
squares the receiving ones. The size of the symbols is proportional to the
ratio of the number of ERASMUS students to the total number of ERASMUS
students in the $30$ academic institutions. The overall fraction of hosted
students in each country is indicated by the intensity of the color of the
country. When the ERASMUS participants are split into two groups (social
sciences and sciences), the female and male mobility patterns are
consistent.~\label{fig::map_network}}
\end{figure}

From Table~\ref{tab::studymob_countries} one concludes that, for the top
sending and receiving countries, female students are systematically
over-represented. But, are there geographical regions that are
preferred by female students more than by their male fellows? The maps in
Fig~\ref{fig::map_network_all} show the geographical distribution of the top
$30$ academic institutions ranked by the number of outgoing and incoming (a)
female and (b) male students, revealing gender differences in the mobility
pattern. Scandinavian universities are definitely more attractive to male
students than to female ones. To understand this effect we analyze the
mobility pattern of ERASMUS participants in the \textit{social
sciences, business and law} and \textit{natural sciences, mathematics and
computing}. For simplicity, we refer to them as \textit{social
science} and \textit{science} groups, respectively. The social science group
consists of $50496$ female and $32011$ male students. The science group is more
balanced, with $7335$ female and $8603$ male students.
Fig~\ref{fig::map_network} shows the geographical distribution of the two
different groups. Within the same subject area there are no significant gender
differences. However, the patterns are significantly different between the two
subject areas. This suggests that the observed gender differences in the
geographical distribution of the top ranked institutions are rather driven by
subject and not by gender. 

\newpage

\section*{Discussion}
The analysis of the mobility in the 2011-12 edition of the ERASMUS program
reveals that female students tend to be over-represented, when compared to
their participation in tertiary education. This over-representation is largely
consistent across subject areas and countries. The study of the geographical
distribution of home- and host-institutions also hints at a gender asymmetry,
suggesting that Scandinavian institutions are more attractive to male students
than to female ones. However, a more detailed analysis shows that the
geographical asymmetry is driven by subject area and consistent with the
distribution of gender ratios among subject areas. 

In the present a study we aim to analyze the existing data without assuming 
any previous postulates. 
This study raises several social questions. What is the reason for this
interesting gender gap in ERASMUS participation? Further studies are necessary. 
One direction for future work might be to investigate 
how social connections among participants affect their choice for the host-institution. For
example, are friends applying for the same university to travel together? Could this be
the mechanism underlying the geographical asymmetry? Also, empirical studies of the
labor market suggest the opposite, namely that female workers are less mobile
than their male partners~\cite{theodossiou09}, a gender gap that even
increases for less-educated workers. The reason for this inversion is still
elusive.  It is also noteworthy that students from \textit{sciences,
mathematics and computing} go to Scandinavia more than they do for Spain and
Italy together. This is in sharp contrast to the agglomerated data, which
suggests that Spain and Italy are very popular countries.

\section*{Supporting Information}

\textbf{S1 File. ERASMUS data.} Data set of ERASMUS participants.

\section*{Acknowledgments}
We would like to thank Olivia Woolley-Meza for fruitful discussions and
acknowledge financial support from the ETH Risk Center, the Brazilian
Institute INCT-SC, ERC Advanced grants numbers FP7-319968 and
FP7-3242247 of the European Research Council, and the Portuguese
Foundation for Science and Technology (FCT) under contract no.
IF/00255/2013 and UID/FIS/00618/2013.  Geographical shape data for
mobility pattern maps were taken from the NUTS 2010 shapefile
\copyright~EuroGeographics for the administrative boundaries.

\bibliography{erasmus}
\bibliographystyle{naturemag}

\pagebreak

\end{document}